\def\etal{{\it et al.\thinspace}}
\def\eg{{\it e.g.\ }}

\def\ie{{\it i.e.\ }}

\def\gsim{~\rlap{$>$}{\lower 1.0ex\hbox{$\sim$}}}
\def\lsim{~\rlap{$<$}{\lower 1.0ex\hbox{$\sim$}}}
  
\documentclass[12pt,preprint]{aastex}

\begin{document}

\title{Behavior of Apsidal Orientations in Planetary Systems}

\author{Rory Barnes\altaffilmark{1} and Richard Greenberg\altaffilmark{1}}

\altaffiltext{1}{Lunar and Planetary Laboratory, University of Arizona,
Tucson, AZ 85721}


\begin{abstract}

A widely considered characteristic of extra-solar planetary systems
has been a seeming tendency for major axes of adjacent orbits to
librate in stable configurations.  Based on a new catalog of
extra-solar planets (Butler \etal 2006) and our numerical
integrations, we find that such small amplitude oscillations are
actually not common, but in fact quite rare; most pairs of planets' major axes are consistenet with circulating relative to one another. However, the new results are
consistent with studies that find that two-planet systems tend to lie
near a separatrix between libration and circulation. Similarly, in systems of
more than two planets, many adjacent orbits lie near a separatrix that divides
modes of circulation.

\end{abstract}

\section{Introduction}

As discoveries and characterizations of extra-solar planetary systems
accumulated, a consensus emerged that the
systems favored alignment of major axes of adjacent orbits about a
fixed orientation (Rivera \& Lissauer 2000; Chiang, Tabachnik \&
Tremaine 2001; Zhou \& Sun 2003; Ji \etal 2003; Go\'zdziewski \&
Maciejewski 2003; Malhotra 2002; Namouni 2005). In 2003
it appeared that 6 out of the 10 then-known systems contained planets
oscillating about such alignments (Zhou \& Sun 2003 based on
Go\'zdziewski \& Maciejewski 2001, 2003; Lee \& Peale 2002, 2003;
Chiang, Tabachnik \& Tremaine 2001; Ji \etal 2003).  In these systems
the difference between the longitudes of periastron of adjacent orbits
librated about either zero (aligned major axes), $180^o$ (anti-aligned
major-axes) or, in the case of 55 Cnc b-c, about $250^o$ (Ji \etal
2003).  Some authors suggested that the apparent prevalence of such
alignments among extra-solar systems may have been important for the
dynamical stability of the systems themselves (Chiang, Tabachnik \&
Murray 2001; Lee \& Peale 2003; Ji \etal 2003; Zhou \& Sun 2003).

With such a large fraction of planetary systems exhibiting this type
of behavior, considerable attention was devoted to explanations of its
occurrence (\ie Malhotra 2002; Chiang, Tabachnik \& Murray 2001; Ford,
Lystad \& Rasio 2005; Namouni 2005).  More recently (with updated
orbits) Barnes \& Greenberg (2006, hereafter BG06) noted that a surprisingly large
fraction of systems could lie near a boundary (here called the
``secular separatrix'') between libration and circulation, that is
they either librate near the maximum possible amplitude, or circulate
with trajectories very close to libration.

A recent catalog of extra-solar systems (Butler \etal 2006) offers the
opportunity to test and update these trends. Based on this catalog, we
have used $N$-body integrations to re-evaluate the behavior of apsidal
motion for all known systems.  In contrast to earlier impressions, we
now find that librating systems may be rare.  However, results that
planets tend to lie near a secular separatrix are reinforced. As with
all current research on extra-solar planets, the results presented
here should be regarded with caution. We use the best-fit orbits from
the Butler \etal catalog (except when the data resulted in an unstable system) which
are certain to change. However, the results presented here are
different from previous published results (\ie Zhou \& Sun
2003). Although these results could be similarly revised in a few
years, they represent the apsidal behavior of planetary systems corresponding to the best current observational data.

\section{Types of Apsidal Motion}
Classical celestial mechanics allows various types of apsidal behavior
for adjacent planet pairs: aligned libration, anti-aligned libration,
non-symmetric libration, circulation, near-separatrix motion between
circulation and libration (see Murray \& Dermott and BG06 for reviews), or near-separatrix motion between modes
of circulation (described below). These types of behavior can be
identified by monitoring $e_1$, $e_2$, and
$\Delta\varpi$ as a function of time, where $e_1$ and $e_2$ are the
eccentricities of the inner and outer planet in a pair, respectively,
and $\Delta\varpi$ is the difference in their longitudes of
periastron.

To first order, analytic determinations of the behavior are only
accurate for small eccentricities (BG06). Hence,
we use numerical integrations with the code HNBODY\footnote{Publicly
available at http://www.astro.umd.edu/$\sim$rauch} (Rauch \& Hamilton
2002), which integrates orbits symplectically, including general
relativity but not tidal dissipation. We will also ignore effects due
to the oblatenesses of the star and planets as there are no
constraints on these parameters. All integrations conserve energy to
better than 1 part in $10^5$, sufficient for this type of integrations
(Barnes \& Quinn 2004). We determine the apsidal behavior by
monitoring the orbital elements over an integration time that covers a
few cycles of $e$ and $\Delta\varpi$ oscillations. For systems
consisting of just a resonant pair we integrated for $10^4$ years,
otherwise we integrated for $10^5$ years (except for the Solar System
which requires $10^6$ years due to the long secular timescale).

The determination of the apsidal mode is generally evident from
consideration of the behavior of $\Delta\varpi$, which can be seen from
a plot of its value vs. time, or from a polar plot of $e_1e_2$ vs.\ $\Delta\varpi$.  In the latter case, if the trajectory encloses the origin, the
system is circulating, otherwise it is librating (e.g.\ BG06).  If the polar trajectory passes close to the origin,
the behavior is near the boundary (i.e.\ a ``separatrix'') between
circulation and libration.  Here we designate such a separatrix between secular libration and circulation as an LCS (libration-circulation separatrix). 

If a planet pair is near a separatrix, the eccentricity of one
of them periodically passes near zero, (\ie BG06, Fig.\ 10, see also Malhotra 2002; Ford,
Lystad \& Rasio 2005). BG06 used $e \le
0.025$ to define near-LCS motion.  Rather than an arbitrary cut-of value for $e$, it may be more
reasonable to compare the minimum value of $e$ to its
range of variation over each cycle. Hence, in this study, a system's proximity to a separatrix is quantified using
\begin{equation}
\label{eq:epsilon}
\epsilon \sim \frac{\textrm{closest approach to origin}}{\textrm{scale of trajectory}} \equiv \frac{2[\textrm{min}(\sqrt{x^2 + y^2})]}{(x_{max} -
x_{min}) + (y_{max} - y_{min})},
\end{equation}
where $x$ and $y$ are the Cartesian coordinates in the polar plot: $x
\equiv e_1e_2\sin (\Delta\varpi)$; $y \equiv e_1e_2\cos (\Delta\varpi)$.
If $\epsilon$ is small (less than a critical value $\epsilon_{crit}$),
then the system lies near some type of separatrix. We will consider
results for two definitions of ``small'', $\epsilon_{crit} = 0.01$ and
0.1, and show that the results are not strongly dependent upon our
choice for $\epsilon_{crit}$. Different estimates for the denominator give similar results. However, for resonant systems this
definition is not meaningful due to the more complicated motion of
this type of interaction (see below).

\begin{figure}
\plotone{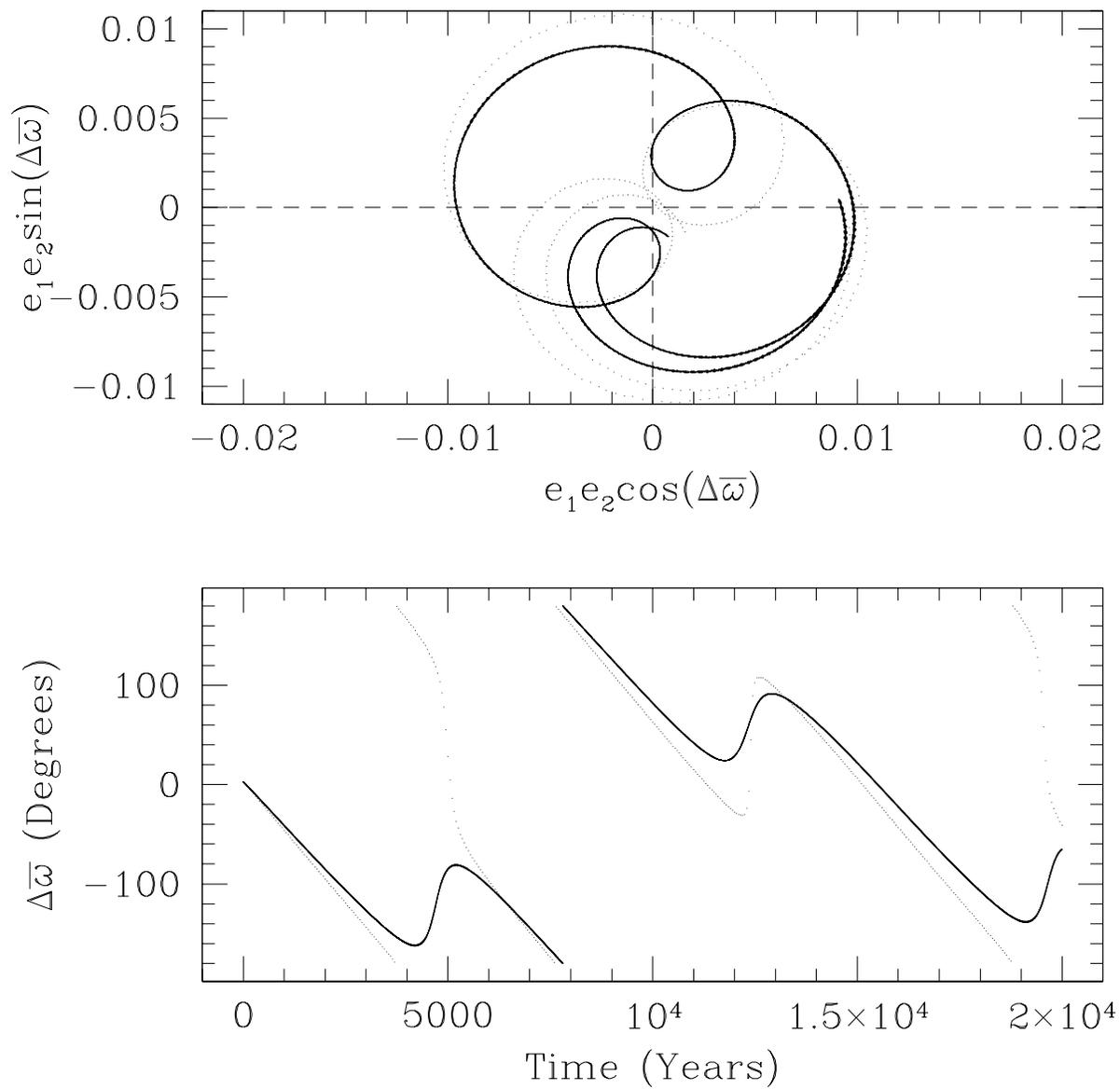}
\figcaption[]{\label{fig:hd69830} \small{Secular evolution of HD 69830 c-d (solid lines), and a system with the eccentricity of the inner planet increased by 50\% (dotted lines). The solid line is the best-fit to the observed system, and the dotted line is a slightly altered, hypothetical system. The solid trajectory and the dotted trajectory are on opposite side of the separatrix (which would pass through the origin in the top panel) that separates two qualitatively different modes of circulation.}}
\end{figure}

In the cases of three or more planets, $\epsilon \le
\epsilon_{crit}$ does not necessarily mean the system is near an
LCS. The secular interaction among all the planets results in a
superposition of oscillations with different amplitudes and different
frequencies (depending on masses, separations, and eccentricities),
which represent eigenmodes of the system (Murray \& Dermott 1999). For
example, in the 3-planet HD 69830 system, consider the mutual apsidal
behavior of planets c and d (Fig.\ \ref{fig:hd69830}). The solid lines
represent the best-fit orbits for this system. Whereas a system with
only 2 planets tends to form an ellipse in $x-y$ space (see BG06 for examples), HD 69830 shows 2 oscillations
superimposed (Fig.\ \ref{fig:hd69830}). In $x-y$ space (top panel),
one oscillation is the large circle that encompasses the origin, the
other appears as an epicycle superposed on the larger circle. These
modes can be identified in the bottom panel ($\Delta\varpi$ vs. $t$)
as well. The first mode corresponds to the nearly linear decrease in
$\Delta\varpi$, whereas the epicycle (in $x, y$) is the small
amplitude oscillation on top of the linear trend.

In a case of only 2 planets, a trajectory that crosses $e=0$ marks the
separation between circulation and libration. For the case of 3
planets, a slight modification of the initial conditions (dotted line)
passes on the opposite side of the origin in the top panel of Fig.\
\ref{fig:hd69830}. However, instead of librating, the new trajectory
is also circulating, but with a different qualitative character
(compare the solid and dotted trajectories in the bottom panel). We
call the separatrix between two types of circulating trajectories a
circulation mode separatrix (CMS). We are unaware of any discussion of
this type of separatrix in the literature.

Our procedure for determining near-separatrix motion is as follows. We
integrate all the planets in a system with HNBODY and calculate
$\epsilon$. We use the minimum masses and coplanar orbits.  If there
are two planets, not in a mean motion resonance, and $\epsilon \le
\epsilon_{crit}$, then the pair is near-LCS. If a system contains three or
more planets and $\epsilon \le \epsilon_{crit}$, then we must
determine the type of separatrix by examining the motion in $x-y$
space. 

For resonant pairs the $x-y$ trajectory can be more complicated and
the generalities above may not apply. Therefore we examined each case
to identify any possible separatrix. Two cases need special
consideration. HD 108874 (Fig.\
\ref{fig:hd108874}) has complex behavior, but clearly switches between apsidal libration and
circulation so it must be near the separatrix. The case of HD 128311
(Fig.\ \ref{fig:hd128311}) is problematic because the
$e-\Delta\varpi$ trajectory does not follow a regular pattern,
presumably due to a complex interplay between the resonant and secular
dynamics. We categorize this
example as circulation of $\Delta\varpi$. Although it might arguably
be near a transition to libration-like behavior. To be conservative,
we do not classify this case as near-separatrix.

\section{Distribution of Apsidal Behavior}
Table 1 lists the results of our simulations and compares them with
earlier statistics. In this table, the letters under ``Mode'' stand
for circulation (C) and libration (L), with a subscript that
corresponds to the equilibrium angle about which $\Delta\varpi$
librates. Where two letters are listed, separated by a slash, the pair
lies near the separatrix between the two modes, with the first letter
corresponding to the mode based on the best-fit orbit. (Note that the
eccentricities of 47 UMa c and GJ 876 d are nominally zero, hence
pairs involving these two planets are precisely on a separatrix by
definition.) Some orbits are poorly constrained (47 UMa c and HD 37124
c), but as we are only conducting a survey, we will include them with
our statistics. C/C means the pair is circulating but near a CMS. We
tabulate the history of best determinations of the orbits in 2003 (from Zhou \& Sun
2003), in 2005 (from Greenberg \& Barnes 2005; BG06), and in 2006 (from this study). The ``Class'' is a general
descriptor of the dynamical character of the system: T represents
systems where one planet is likely to have had its orbit circularized
by tides; R indicates pairs that are in mean motion resonance; U
indicates systems that would be unstable (short-lived) given the
best-fit orbits; S represents all other cases (that is, dominated by
secular interactions).  We also list $\epsilon$ from Eq.\
(\ref{eq:epsilon}) and the minimum eccentricities ($e_1^{min}$ and
$e_2^{min}$) for the inner and outer planets.  

In all these cases we assume best-fit values and do
not compute the errors in $\epsilon$, which must be computed
numerically. To calculate the error we would need to run an exhaustive
suite of simulations varying the orbital parameters within their uncertainties similar to Barnes \& Quinn (2004). Such a survey would be
instructive, but it is beyond the scope of this letter.

In 2003, 60\% of non-tidally-circularized pairs were thought to
oscillate about a specific orientation of the major axes. (We ignore
the tidally evolved systems (T class) because their orbital
history has been controlled in a different way from the other cases,
including both tides and strong relativistic effects because of the
small orbits.) As shown in the 2006 columns, only two (11\%) are
clearly librating given the best-fit orbits. It appears the earlier (and
widely shared) belief that apsidal libration was common was a result
of observational uncertainty in orbital elements.

Although the seemingly strong prevalence of apsidal libration of extra-solar planets (in
2003) now appears not to be the case, the prevalence of near-separatrix
cases noted in 2005 (Greenberg \& Barnes 2005; BG06) is retained. In 2005, 5 out of 16 extra-solar pairs (ignoring unstable and tidally evolved examples)
were near an LCS. Now it appears that 6 out of 18 are near this
boundary, based on the stringent requirement that $\epsilon \le 0.01$
(except for HD 108874 which is near-LCS due to its resonant
interaction). If we include near-CMS behavior then the fraction
increases to 7 of 18.
If we relaxed our requirement for near-separatrix motion to $\epsilon
\lsim 0.1$ (which is not unreasonable) and include T-type pairs then
15 of 24 pairs (over half) would be labeled as near-separatrix.

BG06 also showed that, without systematic effects, the
likelihood of near-separatrix behavior was just a few per
cent. Although observational uncertainties are still relatively large,
the frequency of near-separatrix pairs is larger than expected, given
a uniform distribution of eccentricities. For the T-type pairs, with
damped eccentricities, the fraction of pairs near a separatrix may be
less surprising.

Table 1 also shows that near-LCS motion occurs about as frequently in
R-type systems as S-type systems. Currently 3-4 of 7 R-type
pairs are near an LCS, and 3-4 of 11 S-type pairs (excluding the
Solar System) are near the LCS. Given the small number of pairs, these
proportions are statistically indistinguishable.

We note that two of the tidally-circularized pairs have surprising
characteristics. Planet HD 217107 b has an unusually large
eccentricity for a tidally circularized planet (Vogt \etal
2005). Additionally HD 190360 b is classed as tidally evolved (T)
because it has a small (presumably tidally damped)
eccentricity. However, its period of 17 days is unusually long in
T-type cases; i.e.\ it so far from the star that one would expect
less tidal effect. Nonetheless, we include both these in our T
class. We note that 60\% of T-type systems are near a separatrix, a
fraction similar to that for R- or S-type systems.

\section{Conclusions}
Contrary to general beliefs based on earlier data
(Zhou \& Sun 2003), most extra-solar planetary pairs are probably not in a state of
apsidal libration. Apparently, much more common are systems that lie near a secular separatrix. In addition to near-LCS systems, we find that near-CMS pairs are also unexpectedly common.
The frequency of near-separatrix planets appears to be independent of mean
motion resonances, and perhaps tidal circularization.

As observational data have improved,
systems have tended to be reclassified as near-separatrix. In 2003, 6 out of 10 known systems were thought to be in libration (\eg Zhou \& Sun 2003). By 2005, with 12 known pairs, only 2 librated, while 5 were near-LCS (Greenberg \& Barnes 2005, BG06). Now, of 18 pairs, using the criterion $\epsilon_{crit}\sim 0.1$, we find 11 near-separatrix (of which 9 are near-LCS), while only 2 librate. Even with the more stringent definition of ``near'' ($\epsilon_{crit} = 0.01$), we find 7 of 18 are near a separatrix (of which 6 are near-LCS).

The previous notion that most systems librated was due to large
uncertainties in orbits. The results presented here may be revised as
the orbital data continue to improve and the distribution of apsidal
behavior presented here will be refined as more systems are
discovered. Given the fraction
of pairs that is near-separatrix (even for $\epsilon_{crit} = 0.01$), it
is unlikely that improved orbit determinations will change  the results for the known cases. The $\upsilon$ And c-d pair
has a relatively small error ellipse, and the likelihood that this
system is far from the separatrix is quite small (Ford, Lystad \&
Rasio 2005).  Furthermore two of the three gas-giant pairs in our
Solar System (which suffer no appreciable observational errors) are
near a CMS.

Regardless of observational biases, the current catalog of extra-solar
planets contains a significant number of near-LCS
planets. Given the earlier belief that libration was common, the fraction of near-LCS systems is surprising. Assuming this result is robust, how might it be explained? Planet-planet scattering has been suggested in one case (Ford, Lystad \& Rasio 2005), but it is
unknown if this phenomenon produces near-CMS motion or the observed
distribution of near-LCS motion. Moreover, if planet-planet scattering
is the cause, then the nature of near-LCS motion in R-type systems
must be explained. Most models for the formation of R-type systems are
based on adiabatic resonance capture during migration in a gaseous
protoplanetary disk (Kley, Peitz \& Bryden 2004). This type of
interaction is fundamentally different from the impulsive process of
scattering. The recent work of S\'andor \& Kley (2006) showed that a combination of migration and scattering may produce near-separatrix motion in resonant systems. This hypothesis deserves further consideration in the context of apsidal behavior.

\begin{figure}
\plotone{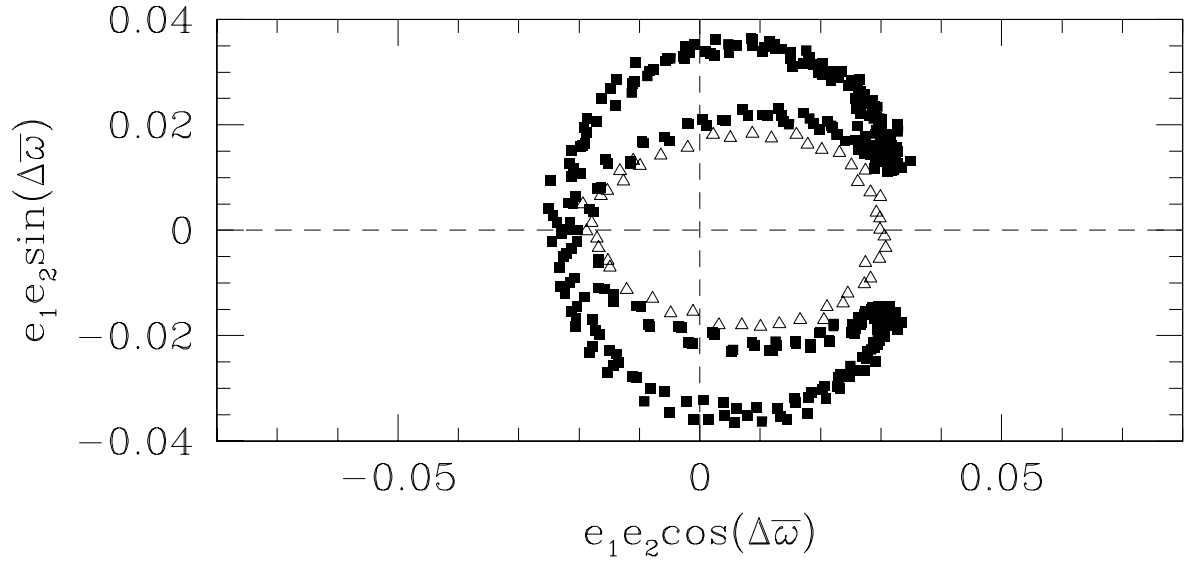}
\figcaption[]{\label{fig:hd108874} \small{Apsidal trajectory of the resonant system HD 108874. The open triangles are data from the first 2500 years of evolution of HD 108874, and the filled squares are from 50,000 -- 67,000 years. Initially the pair circulates, but at later times it librates with high amplitude. This pattern repeats on a $\sim 10^5$ year timescale.}}
\end{figure}

\begin{figure}
\plotone{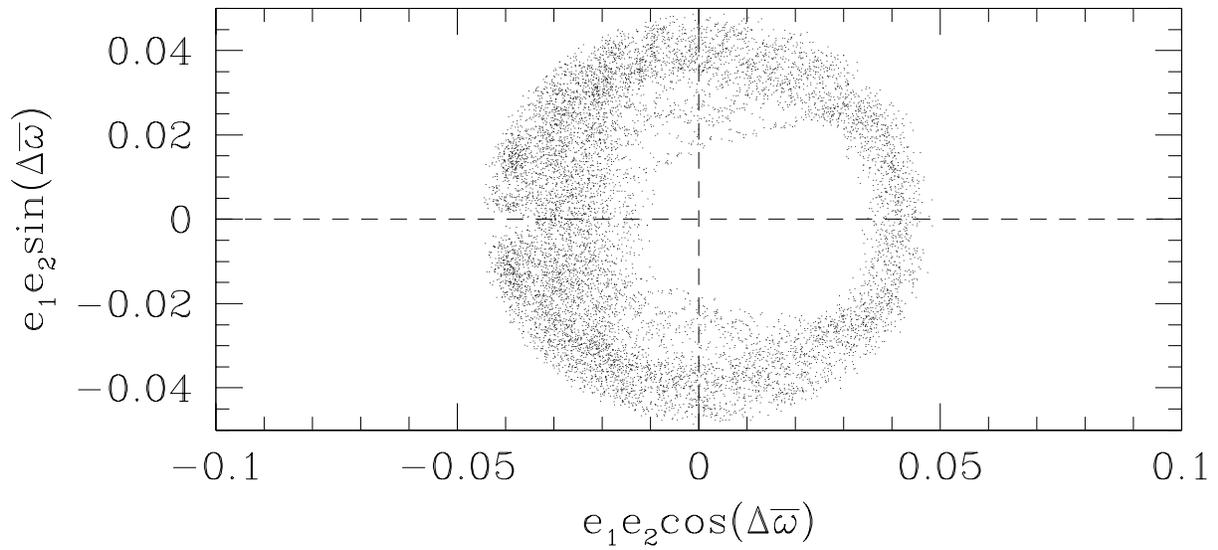}
\figcaption[]{\label{fig:hd128311} \small{Apsidal trajectory of the resonant system HD 128311. The points represent the system at 1 year intervals for 10,000 years. HD 128311 passes close to the origin ($\epsilon = 0.09$), but the complexity of the trajectory (due to the resonant interaction) makes this an unusual system, not readily classified in terms of libration or circulation.}}
\end{figure}

Future work should also find a more suitable definition of
``near-separatrix'' that encompasses resonant pairs. The complexities of resonant dynamics made a universal definition beyond the scope of this letter.  The true behavior is simple enough to determine
from an examination of the motion (Figs.\ \ref{fig:hd108874}--\ref{fig:hd128311}), but a fast, accurate method to
determine near-separatrix motion is needed.

The current secular interactions of planets may bear the
imprint of past dynamical events (Malhotra 2002; Ford, Lystad \& Rasio
2005). The observed distribution of apsidal
modes described here may help constrain planet formation models (\ie
Boss 2002; Mayer \etal 2002). Future observations will test this
characteristic of planetary systems.

\medskip
This work was funded by NASA's Planetary Geology and Geophysics
program grant number NNG05GH65G. We would also like to thank two anonymous referees for suggestions that greatly improved this manuscript.

\references
Barnes, R. \& Greenberg 2006, ApJ, 638, 478 (BG06)\\
Barnes, R. \& Quinn, T.R. 2004, ApJ, 611, 494\\
Boss, A.P. 2001, ApJ, 563, 367\\
Butler, R.P. \etal 2006, ApJ, 646, 505\\
Chiang, E. I., Tabachnik, S., \& Tremaine, S. 2001, AJ, 122, 1607\\
Correia, A.C.M. \etal 2005, A\&A, 440, 751\\
Ford, E.B., Lystad, V., \& Rasio, F.A. 2005, Nature, 434, 873\\
Kley, W., Peitz, J., \& Bryden, G. 2004, A\&A, 414, 735\\
Go\'zdziewski, K. \& Maciejewski, A. 2001, ApJL, 563, L81\\
------------. 2003, ApJL, 586, L153\\
Greenberg, R. \& Barnes, R. 2005, BAAS, 37, 672\\
Ji, J. H., Kinoshita, H., Liu, L., \& Li, G. Y. 2003, ApJ, 585, L139\\
Lee, M. H., \& Peale, S. J. 2002, ApJ, 567, 596\\
------------. 2003, ApJ, 592, 1201 (erratum, 597, 644)\\
Malhotra, R. 2002, ApJL, L33\\ 
Mayer, L. \etal 2002, Science, 298, 1756\\
Murray, C.D. \& Dermott, S.F. 1999 \textit{Solar System Dynamics}, Cambridge UP, Cambridge\\
Namouni, F. 2005, AJ, 130, 280\\
Laughlin, G., Chambers, J., \& Fisher, D. 2002, 579, 455\\
Lovis, C. \etal 2006, Nature, 441, 305\\
Rasio, F. \& Ford, E. 1996 Science, 274, 954\\
Rauch \& Hamilton 2002, BAAS, 34, 938\\
Rivera, E.J. \& Lissauer, J.J. 2000, ApJ, 530, 454\\
Tinney, C.G. \etal 2006, ApJ, 647, 594\\
Vogt, S.S. \etal 2005, ApJ, 632,638\\
Zhou, J.L. \& Sun Y.S., 2003, ApJ, 598, 1220\\

\clearpage
\clearpage

\clearpage
\clearpage

\clearpage
\clearpage

\begin{center}Table 1: Apsidal Motions of Multi-planet Systems\end{center}
{\small
\begin{tabular}{cc|cccc|cccc}
\hline\hline
 & & & & Mode & & & & & \\
\hline
System & Pair & 2003 & 2005 & 2006$^a$ & 2006$b$ & $\epsilon$ & $e_1^{min}$ & $e_2^{min}$ & Class\\
\hline
SS$^c$ & J-S & C & C & C & C & 0.194 & 0.016 & 0.017 & S\\
   & S-U & C & C & C/C & C/C & 0.006 & 0.017 & $9\times 10^{-4}$ & S\\
   & U-N & C & C & C/C & C/C & 0.004 & $9.3\times 10^{-4}$ & $4.4\times 10^{-4}$ & S\\
55 Cnc$^d$ & e-b & - & C & C & C/C & 0.067 & 0.061 & 0.0045 & T\\
   & b-c & L$_{250}$ & C/L$_{250}$ & C & C/L$_{180}$ & 0.110 & 0.0045 & 0.033 & R\\
   & c-d & C & C & C & C & 0.158 & 0.033 & 0.079 & S\\
$\upsilon$ And$^d$ & b-c & C & C & C/C & C/C & $1.8\times 10^{-4}$ & 0.014 & $8.3\times 10^{-5}$ & T\\
   & c-d & L$_0$ & L$_0$/C & C/L$_0$ & C/L$_0$& $2.8 \times 10^{-4}$ & $8.3\times 10^{-5}$ & 0.23 & S\\
HD160691$^d$ & d-b & - & - & - & - & - & - & - & U\\
   & c-b & - & - & - & - & - & - & - & U\\
GJ876$^d$ & d-c & - & - & C/C$^h$ & C/C$^h$ & 0 & 0 & 0.22 & T\\
   & c-b & L$_0$ & C/L$_0$ & L$_0$ & L$_0$ & 0.34 & 0.22 & 0.013 & R\\
HD 69830$^e$ & b-c & - & - & C & C/L$_{180}$ & 0.095 & 0.06 & 0.011 & T\\
   & c-d & - & - & C & C/C & 0.040 & 0.011 & 0.069 & S\\
HD 37124$^d$ & b-c & - & - & C/C & C/C & 0.009 & 0.024 & 0.021 & S\\
   & c-d & - & - & L$_0$ & L$_0$/C & 0.096 & 0.021 & 0.09 & S\\
47 UMa$^d$ & b-c & L$_0$ & C/L$_0$ & C/L$_0$$^i$ & C/L$_0$$^i$ & 0 & 0.02 & 0 & S\\
HD 12661$^d$ & b-c & L$_{180}$ & L$_{180}$ & C/L$_{180}$ & C/L$_{180}$ & 0.003 & 0.09 & $6\times 10^{-4}$ & S\\
HD 38529$^d$ & b-c & C & C & C & C & 0.44 & 0.23 & 0.33 & S\\
HD 73526$^f$ & b-c & - & - & L$_{180}$/C & L$_{180}$/C & 0.006 & 0.15 & 0.0008 & R\\
HD 74156$^d$ & b-c & C & C & C & C & 0.36 & 0.43 & 0.56 & S\\
HD 82943$^d$ & b-c & L$_0$ & U & C/L$_0$ & C/L$_0$ & 0.004 & 0.080 & 0.001 & R\\
HD 108874$^d$ & b-c & - & L$_{180}$ & C/L$_{180}$$^j$ & C/L$_{180}$$^j$ & 0.198 & 0.05 & 0.074 & R\\
HD 128311$^d$ & b-c & - & L$_0$/C & C & C$^k$ & 0.091 & 0.06 & 0.03 & R\\
HD 168443$^d$ & b-c & C & C & C & C & 0.219 & 0.48 & 0.15 & S\\
HD 169830$^d$ & b-c & - & C & C & C & 0.326 & 0.19 & 0.21 & S\\
HD 190360$^d$ & c-b & - & C & C & C & 0.38 & 0.009 & 0.36 & T\\
HD 202206$^g$ & b-c & - & C & C & C/L$_0$ & 0.096 & 0.41 & 0.064 & R\\
HD 217107$^d$ & b-c & - & C & C & C & 0.46 & 0.13 & 0.52 & T\\
\end{tabular}
\\
$^a$ $\epsilon_{crit}\sim 0.01$, $^b$ $\epsilon_{crit}\sim 0.1$\\
$^c$ http://ssd.jpl.nasa.gov, $^d$ Butler \etal (2006), $^e$ Lovis \etal (2006), $^f$ Tinney \etal (2006), $^g$ Correia \etal (2005)\\
$^h$ The mode of this pair is on a separatrix, but the resonant interaction complicates the motion.\\
$^i$ The mode of this pair is on the C/L$_0$ separatrix.\\
$^j$ Although $\epsilon > 0.1$ for this system, the resonant interaction places it close to the LCS.\\
$^k$ Although $\epsilon < 0.1$ for this system, the resonant interaction complicates the motion.\\

\end{document}